\documentclass[11pt]{amsart}

\pdfoutput=1

\usepackage{cite}

   \usepackage[pdftex]{graphicx}
   \graphicspath{{./png/}}
   \DeclareGraphicsExtensions{.pdf,.jpeg,.png,.tiff}

\usepackage{array}

\usepackage{mdwmath}
\usepackage{mdwtab}

\usepackage[caption=false,font=footnotesize]{subfig}

\begin{document}

\title[IEEE Trans Evolutionary Computation 16 (2012) 711--729. ]{Evolution of Plastic Learning in Spiking Networks via Memristive Connections}

\author[Howards, Gale, Bull, De Lacy Costello, Adamatzky]
{Gerard~Howard,
        Ella~Gale,
        Larry~Bull, \\
        Ben~de~Lacy~Costello,
        Andy~Adamatzky
}

\thanks{Gerard Howard, Larry Bull and Andrew Adamatzky are with the department of Computer Science and Creative Technologies, University of the West of England, Bristol BS16 1QY, UK, contact e-mail: david4.howard@uwe.ac.uk.}
\thanks{Ella Gale and Ben de Lacy Costello are with the department of Chemistry, University of the West of England, Bristol BS16 1QY, UK.}


\maketitle

\begin{abstract}
\label{abstract}
 This article presents a spiking neuroevolutionary system which implements memristors as plastic connections, i.e. whose weights can vary during a trial.  The evolutionary design process exploits parameter self-adaptation and variable topologies, allowing the number of neurons, connection weights, and inter-neural connectivity pattern to emerge.  By comparing two phenomenological real-world memristor implementations with networks comprised of (i) linear resistors (ii) constant-valued connections, we demonstrate that this approach allows the evolution of networks of appropriate complexity to emerge whilst exploiting the memristive properties of the connections to reduce learning time.  We extend this approach to allow for heterogeneous mixtures of memristors within the networks; our approach provides an in-depth analysis of network structure.  Our networks are evaluated on simulated robotic navigation tasks; results demonstrate that memristive plasticity enables higher performance than constant-weighted connections in both static and dynamic reward scenarios, and that mixtures of memristive elements provide performance advantages when compared to homogeneous memristive networks. 
 
 
\end{abstract}

\section{Introduction}
\label{intro}

The field concerned with nanoscale brainlike information processing is known as Neuromorphic Computation (NC) \cite{meadneuromorphic}.  NC is a new way of computing in hardware that blurs the distinction between processor and memory, as both may be distributed at any spatial position in the architecture.  Neuron-like units, such as Complementary Metal-Oxide Semiconductor (CMOS) neurons \cite{cmos-overview}, are densely interconnected by numerous adaptive synapses that communicate via transmission of spikes.  Neuromorphic architectures are yet to be physically realised, yet are envisioned to encorporate many attractive characteristics, including redundancy, self-organisation, adaptation, and learning \cite{meadneuromorphic}. 

NC has recently become more viable thanks to the manufacture of the memristor \cite{chua-mem} (memory resistor) at HP labs \cite{missing-mem-found}.  A memristor is a fundamental passive two-terminal circuit element whose state (memristance) is both nonvolative and dependent upon past activity.  Nonvolative memory \cite{nonvol-mem-mem} is perfect for low-power storage, and the devices dynamic internal state facilitates information processing.  These properties make the memristor ideal for use as a nanoscale synapse in NC architectures \cite{bio-inspired-adapt-mem-nets}.  A proposed approach to realise learning in NC involves harnessing Hebbian principles \cite{hebb43} to realise Spike Time Dependent Plasticity (STDP) \cite{stdp},  allowing connections between a presynaptic and postsynaptic neuron to alter efficiacy dependent on the spike timings of those neurons. 

It has been reasoned that, much like the brain, different areas of NC architectures could be responsible for different activities.  To this end, we focus on the evolution of self-organizing small-scale Spiking Neural Networks (SNNs \cite{spiking-n-m}), where each network can be conceptualised as being representative of part of a larger NC architecture.  We employ a model of neuroevolution whereby each network in the population initially comprises of a number of hidden layer neurons, connected to a problem-dependent, fixed number of input and output neurons.  The evolutionary process can then alter network topology as part of the Genetic Algorithm (GA) \cite{holland76}.  

In this study, we initially compare two phenomenological memristor implementations, Hewlitt-Packard (HP)-like \cite{missing-mem-found} and Polyethylene Oxide - Polyaniline (PEO-PANI)-like \cite{peo-0}, and analyse their computational properties when cast as synaptic connections in evolutionary SNNs.  The memristive element of the network is designed to allow the weight of the connections to vary during a trial, providing a learning architecture which may be beneficial to the evolutionary design process.  We then allow for the evolution of heterogeneous memristive networks (e.g. those containing all memristor types), and investigate whether such mixtures give an inherent performance advantage when compared to their homogeneous counterparts.  As the equations used to govern the memristors are based on physical devices, the evolved networks represent possible behaviours of partial NC architectures (e.g. in the context of evolvable hardware \cite{788492}).  Performance is evaluated on simulated robotic navigation tasks.   

Our initial hypothesis is that memristive synapses provide the networks with increased performance.  To test this hypothesis, we compare the homogenous memristive networks (HP, and PEO-PANI) to networks solely comprised of linear resistors (e.g. \cite{662327}) and constant-weighted elements.  Extending the hypothesis to heterogeneous networks, we seek to confirm that varied memristive behaviours can be harnessed by the evolutionary design process to provide further advantages, specifically that (i) certain functionality can be more easily achieved by certain memristor types (ii) combinations of memristor types are beneficial to the networks.  Specifically, we aim to answer the following research questions:\\

\begin{enumerate}

\item Does the evolutionary process allow for the successful generation of memristive networks that outperform constant-valued connections,  despite the memristors nonlinearity and given the potential for complex interactions within memristive networks?

\item In the heterogeneous case, do mixtures of memristors provide better performance than other implementations?   How do such networks generate useful behaviour?

\item Is there an evolutionary preference in assigning specific roles to specific memristor types based on variations in their memristive behaviours?

\end{enumerate}

\subsection{Roadmap}
\label{roadmap}
The remainder of the article is ordered as follows: Section II introduces background research.  Section III introduces the system. Section IV details the spiking implementation. Section V outlines memristor implementations.  Section VI details the GA.  Section VII details the network topology mechanisms.  Section VIII gives the environment.  Section IX details the experimental setup. Sections X, XI,and XII analyse the results of the experiments that were carried out and highlight the main differences between the memristor models.  Section XIII provides a summary.

\section{Background}
\label{background}

\subsection{Spiking Networks and Evolutionary Spiking Networks}
\label{snn+esnn}
SNNs  present a biologically plausible phenomenological model of neural activity in the brain.  In a SNN,  neurons are linked via unidirectional, weighted connections that act as communication carriers.  When two neurons (A and B) are connected, neuron A is either (a) presynaptic to neuron B (the connection is directed from neuron A to neuron B) or (b) postsynaptic to neuron B, if the connection is directed from neuron B to neuron A.  

The medium of communication is the action potential, or spike, which is emitted from a presynaptic neuron and received by all connected postsynaptic neurons.  Each neuron has an internal state, known as``membrane potential", which is influenced by spike reception but decreases over time.  Spikes are emitted from a given neuron after this state surpasses a certain level of excitation (received either from the environment or from presynaptic neurons).  This time-dependent build-up of membrane potentials and release of spikes is able to produce dynamic activation patterns through time.  

The earliest equations that describe SNNs were described by Lapicque in 1907 \cite{lapicque}.  Two popular formal SNN implementations are the Leaky Integrate and Fire (LIF) model \cite{spiking-n-m} (which is derived from \cite{lapicque}) and the Spike Response Model (SRM) \cite{spiking-n-m}.  The main justifications for including SNN networks are (i) increased utility when compared to other network models e.g. the MLP \cite{rumelhart} - shown in \cite{maass,  sag-wex} (ii) current NC research focussing on spiking neurons as a basis of communication due to low power requirements and the ability to harness spikes as a learning mechanism (e.g \cite{5596372, 4693998}).   

Application of evolutionary techniques to neural networks involves the use of a GA to alter connection weights, network topology, connectivity patterns, or combinations of the above.  A survey of various methods for evolving both weights and architectures in neural networks is presented in \cite{neuroevo-arch-learn}. Neuro-evolution was first applied to LIF SNNs to evolve networks that produce temporally-dependent outputs \cite{korkin-spike-cbm} and SRM spiking networks were first evolved for a vision-based navigation task \cite{evo-snn-vis-based-rob}. 

As the subject of the paper describes robotics tasks, a short overview of spiking neuroevolutionary robotics follows.  Nolfi and Floreano \cite{evo-rob-book}  provide a review.  SNN circuits were evolved to model abstractions of biological retina in \cite{indivieri-neuro-vlsi} where the system was applied to a robotics platform.  An LIF spiking model is used in \cite{evo-bns}, again for the goal of evolving  navigation behaviours.  A similar spiking model is applied to a simple robotic navigation task \cite{evo-developing-snn}; the authors conclude that the dynamics of a SNN provide further degrees of problem-solving freedom given temporally-sensitive problems.  A recent hardware implementation is given in \cite{recentspikerob}.

\subsection{Memristors}
\label{memristors}
Memristors (memory-resistors) are the fourth fundamental circuit element, joining the capacitor, inductor and resistor \cite{insights-into-memristor}.  A memristor can be defined as a resistor whose instantaneous resistance value (a) depends on all charge that has passed through it (b) is nonvolatile.  Formally, a memristor is a passive two-terminal electronic device that is described by the non-linear relationship between the device terminal voltage, $v$, and terminal current, $i$, as shown in (1).  Nonlinearity arises because the instantaneous memristance, $M$, depends on the charge $q$ (2), where $\varphi$ is the time integral of voltage, or magnetic flux.\\

$v = M(q)i$                   \hfill (1)\\

$M(q) = d\varphi(q) / dq $        \hfill(2)\\

The memristor was theorectically characterized and named by Chua in 1971 \cite{chua-mem}.  Memristive systems have recently enjoyed a resurgence of interest from the research community after being manufactured by HP labs \cite{missing-mem-found}.  This has spawned a number of research avenues in terms of applications of memristive systems \cite{mem-based-multilevel-memory}\cite{mem-pattern-recognition}, and synthesis of various other memristors \cite{macro-memristor} \cite{nano-mem-syn-neuromorphic}.

There are many reasons to think that memristors might be useful in NC.  Primarily, memristors can be manufactured at the required scale and implement synaptic behaviour in hardware \cite{linares-barranco}.  HP memristors \cite{missing-mem-found} have been used in the manufacture of nanoscale neural crossbars \cite{comp-resis-xbar}, which have been applied to pattern recognition circuits \cite{mem-pattern-recognition}.  Silver Silicide memristors have  been shown to function in neural architectures \cite{nano-mem-syn-neuromorphic}. Memristor theory has also been used to model learning in amoeba \cite{mem-amoeba}.  In particular, \cite{mem-pattern-recognition} highlights the attractive prospect of applying evolutionary computation techniques directly to memristive hardware, as memristors can simultaneously perform the functions of both processor and memory.  This work will focus on the use of the memristor as an adaptive synapse.

\subsection{Synaptic Plasticity}
\label{plasticity}

Hebbian learning \cite{hebb43} is thought to account for synaptic adaptation and learning in the brain.  Briefly, Hebbian learning states that ``neurons that fire together, wire together'' - in other words in the event that a presynaptic neuron causes a postsynaptic neuron to fire, the synaptic strength between those two neurons is increased so that such an event is more likely to happen in the future.  This mechanism allows for self-organising, correlated activation patterns and is therefore of particular relevence when considering learning in neural systems.  

Spike Time dependent Plasticity (STDP) \cite{stdp} was originally formulated as a way of implementing Hebbian learning within computer-based neural networks.  Interestingly, the STDP equation has been found to have distinct similararities to the reality of Hebbian learning in biological synapses \cite{bi-poo}.  It has recently been postulated that a memristance-like mechanism affects synaptic efficiacy in biological neural networks \cite{linares-barranco}, based on similarities between memristive equations and their neural counterparts.

Integration of neuroevolution with neuromodulatory networks is investigated by Soltoggio (e.g., \cite{soltoggio-het}).   In the networks, dedicated modulatory neurons are responsible for affecting the inputs received by traditional neurons.  Heterogeneous modulation rules are available to the networks, although unlike memristors they have no direct hardware analogue.  The networks are tested on agent navigation tasks and robot controllers \cite{neuromod-robot}, both with promising results.  A probabilistic SNN model is investigated \cite{maass-zador} whereby the probability of spike transmission across a synapse is affected by Hebbian learning rules; results demonstrate the power of plasticity in generating varied behaviour.  Floreano and Urzelai \cite{flor-urz} evolve  Discrete Time Recurrent Neural Networks \cite{beer-minimal-cog} where synapses are affected by four versions of the Hebb rule during the lifetime of the agent as it solves a navigation task.

Memristive STDP has been implemented in \cite{stdp-discrete}\cite{nano-mem-syn-neuromorphic}\cite{stdp-nano-cmos-asyn}\cite{linares-barranco}, with all four papers using spike-coincidence based STDP as a learning mechanism.  Also consistent between the papers is the use of a ``two-part spike'', which a SNN neuron uses to pass information to both presynaptic and postsynaptic neurons.  The temporal coincidence between presynaptic and postsynaptic spikes at a memristive synapse alters the voltage across that synapse; if a threshold voltage is surpassed, the synapses weight is altered.  The main difference between \cite{stdp-discrete}\cite{nano-mem-syn-neuromorphic} and \cite{stdp-nano-cmos-asyn}\cite{linares-barranco} is that in \cite{stdp-discrete}\cite{nano-mem-syn-neuromorphic}, the two-part spike is implemented as a discrete-time stepwise waveform approximation, whereas \cite{stdp-nano-cmos-asyn}\cite{linares-barranco} use values calculated from continuous waveform equations, allowing them to operate in continuous time.  

In summary, this literature review has highlighted the previous success of STDP as a neural learning mechanism, and shown memristors as an ideal medium for the implementation of STDP, especially coupled with a SNNs.  The relevence of a robotic navigation task in the context of neuroevolution is also shown.\\\\

\section{The System}
\label{the-system}
The system presented here consists of a population of SNNs which are evaluated on a robotics test problem, and altered via GA operation which is detailed in section VI.  To introduce the terminology to be used throughout this paper: each experiment lasted for 1000 evolutionary {\em generations}; each generation involved new networks in the population being evaluated on the test problem (a {\em trial}).  Each trial consisted of a number of {\em timesteps}, which began with the reading of sensory information and calculation of action, and ended with the agent performing that action.  Every timestep consisted of 21 {\em steps} of SNN processing, at the end of which the action was calculated.

\section{Spiking Network Implementation}
\label{snn-imp}
We base our spiking implementation on the LIF model.  Neurons can be stimulated either by an external current or by connections from presynaptic neurons; recurrency and direct input - output connections are illegal.  Each neuron has a membrane potential, $y$, where $y$$>$0, which slowly degrades over time according to (3).  As spikes are received by the neuron, the value of $y$ is increased in the case of an excitory spike, or decreased if the spike is inhibitory.  If $y$ surpasses a positive threshold, $y_{thresh}$, the neuron spikes and transmits an action potential to every neuron to which it is presynaptic, with strength relative to the efficiacy of the synapse that connects them.  The neuron then resets its membrane potential to some low number.  At time $t$, the membrane potential of a neuron is given in (3); the reset equation is given in (4).\\

$y(t+1) = y(t) + (I+a-by(t))$ \hfill(3)\\

$If (y > y_{thresh}) y = c$ \hfill(4)\\

Here, $y(t)$ is the membrane potential at time $t$, $I$ is the input current to the neuron, $a$ is a positive constant, $b$ is the degradation (leak) constant and $c$ is the reset membrane potential of the neuron.  The networks are arranged into three layers: input (which receives sensory information), hidden (a variable-size layer), and output (where motor-actions are calculated). Example architectures can be seen in Figs. ~\ref{activation-all}(d), ~\ref{fig_sim} and ~\ref{dyn-all}(d).  A model of temporal delays is used so that, in the single hidden layer only, a spike sent between two neurons is received $x$ steps after it is sent (see (5)), $i_s$ is the index of the sending neuron and $i_r$ is the index of the receiving neuron (indexing is sequential).\\

$x = i_s - i_r$\hfill(5)

\subsection{Action Calculation}
\label{snn-act-calc}
Action calculation involves the current input state being repeatedly processed 21 times by each network (experimentally determined to allow sufficient STDP to occur during the lifetime of the agent).  For the purposes of this paper, each network was initialised with 6 input neurons (used to pass sensor values to the network), nine hidden neurons, and 2 output neurons that are used to calculate the action.  Each output neuron had an activation window that recorded the number of spikes produced by that neuron over the last 21 steps. We classified the spike trains at the two output neurons as being either {\em low} or {\em high} activated (see (6)).  \\

$if(n_s < t_s / 2) \{low\}$
$else \{high\}$\hfill(6)\\

Here, $n_s$ is the number of spikes in the window and $t_s$ is the window size.  The combined spike trains at the two output neurons translated to a discrete movement according to the output activation strengths.  See section ~\ref{agent} for precise details of sensory state generation and possible actions.

\section{Variable Connections}
\label{memristive-connections}
We are primarily interested in implementing memristors as a form of variably-weighted connection between neurons in our SNNs, where variable weight indicates that connection efficiacy can alter {\em during} a trial.  The behaviour of each memristor under STDP depends on the memristive equations used; these are defined in subsections 1 and 2.  The linear resistor is described in subsection 3.  It is important to note that as the calculated resistance value of memristive connections are based on their real-world counterparts, simulation results should be replicable in hardware.  Constant-valued connections do not alter weight during a trial; rather, their weights are altered between trials via the GA.  The HP memristor was chosen for study as it is well understood.  The PEO-PANI memristor was chosen as it is also well-understood, but more importantly has a strongly different memristance profile (see Fig.~\ref{stdpfig}(a)), allowing the potential for contrasting behaviour.\\

\subsubsection{HP Memristor}
\label{hp-mem}
The HP memristor is comprised of thin-film Titanium Dioxide (TiO$_{2}$) and oxygen-depleted Titanium Oxide (TiO$_{2-x}$).  The boundary between the two compounds moves in response to the charge on the memristor, which in turn alters the resistance of the device.  More details can be found in \cite{pickett}.  Memristance is defined in (7):\\

$M = R_{off} - R_{off} \times R_{on} \times \beta \times q $ \hfill (7)\\

$R_{off}$ is the resistance of the TiO$_{2}$ and $R_{on}$ is the resistance of the TiO$_{2-x}$.  $\beta$ is a parameter comprising both the thickness of the device and the mobility of the oxygen vacancies in TiO$_{2}$ and TiO$_{2-x}$, and $q$ ($0 \leq q  \leq  q_{max}$) (see (8)) is the charge on the device.\\

$q_{max} = (R_{on} - R_{off}) / - R_{on} \times R_{off} \times \beta$ \hfill(8)\\

One further parameter is $mem\_lifetime$, which is the amount of time it takes the memristor to alter from $M=R_{off}$ to $M=R_{on}$.  Once $M$ is calculated, the weight $W$ of the memristive connection can be set as $1/M$; weight is therefore equal to the inverse memristance of the device.\\

\subsubsection{PEO-PANI Memristor}
\label{peo-mem}
The PEO-PANI memristor consists of layers of PANI, onto which Lithium ion (Li$^+$)-doped PEO is added \cite{peo-0}.  We have phenomenologically recreated the memristance profile of the PEO-PANI memristor, resulting in behaviour similar to that seen in \cite{peo-0}.  The equation for $M$ is identical to that of the HP memristor (7); the weighting equation is given in (9):\\

$W = 1 - (1/R_{off} + R_{on} - M) + (1/R_{off}) $ \hfill(9)\\

\subsubsection{Linear Resistor}
\label{var-mem}
In this study the term ``linear resistor'' refers to a theoretical nonvolatile-memory-augmented device that describes a linear relation between $i$ and $v$.  The linear resistor alters $W$ by $1 / mem\_lifetime$, therefore it takes $mem\_lifetime$ positive STDP events to  increase $W$ from $R_{off}$ to $R_{on}$. 

\begin{figure}
\centering
	\subfloat[]{\label{stdp-full}\includegraphics[width=3.2in]{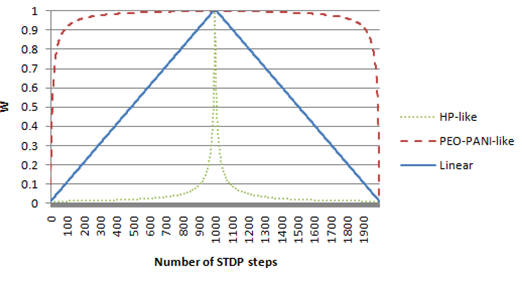}}     \\           
	\subfloat[]{\label{stdp-sens}\includegraphics[width=3.2in]{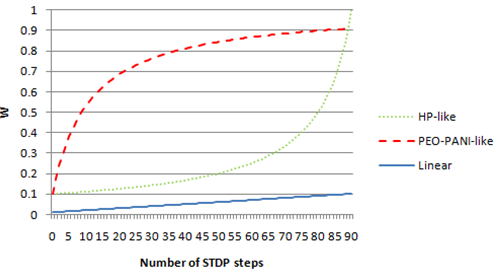}}
\caption{ (a) Full range of memristance for the three variable connection types, given 1000 positive STDP events followed by 1000 negative STDP events, with $mem\_lifetime$=1000 (b) sensitive ranges of memristance curves.  The PEO-PANI-like and linear resistor sensitive ranges are described by STDP steps $0-90$ in Fig. 1(a); the HP-like sensitive range includes STDP steps $910-1000$.}
\label{stdpfig}
\end{figure}

\subsection{STDP}
\label{stdp}

\begin{figure}[!t]
\centering
\includegraphics[width=3in]{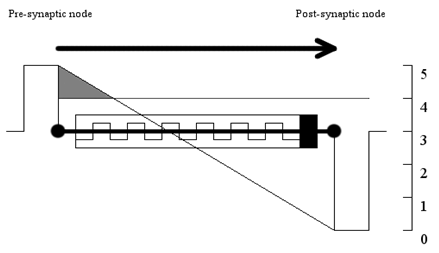}
\caption{Visualising the HP memristor when implemented as a synapse, showing a positive STDP event.  The numerical scale (right) shows the sum of $last\_spike$ events, where the presynaptic node has $last\_spike$=2 and the postsynaptic node has $last\_spike$=3.  Given $spike\_threshold=4$, memristance moves the $TiO_{2}$ - $TiO_{2-x}$ boundary so there is more $TiO_{2}$ than  $TiO_{2-x}$; resistance decreases leading to an increase in synaptic efficiacy.}
\label{figsim}
\end{figure}

As mentioned in section ~\ref{plasticity}, a number of STDP implementations exist.  As our SNNs operated in discrete time, we follow \cite{stdp-discrete}\cite{nano-mem-syn-neuromorphic} in using discrete-time stepwise waveforms.  STDP could affect any variable connection in the networks.

In our implementation, each neuron was augmented with a variable to record the last time it spiked ($ls$), which is initially 0.  When a neuron spiked, $ls$ was set to some positive number.  At the end of each of the 21 steps that make up a single timestep, each memristive connection is analysed by checking the $ls$ values of its presynaptic and postsynaptic neurons.  If the calculated value exceeded a positive threshold $\theta_s$, memristance of the synapse occurred (see Fig.~\ref{figsim},(10),(11)). At the end of each step, each $ls$ value was decreased by 1 to a minimum of 0, creating a discrete stepwise waveform through time. Each STDP event altered $q$, by $\Delta q$, as detailed in (12), which was then used to calculate the synaptic weight.  The memory of the system is therefore contained in $q$.\\

$ if(ls_{pre} + ls_{post} > \theta_s  \,AND\, ls_{pre} > ls_{post}) \{q  \stackrel{-}{=}\Delta q\} $ \hfill(10)\\

$ if(ls_{pre} + ls_{post} > \theta_s  \,AND\, ls_{pre} < ls_{post}) \{q \stackrel{+}{=} \Delta q\} $ \hfill(11)\\

$\Delta q = q_{max} / mem\_lifetime $\hfill(12)\\

From Fig. 1(a) it can be seen that the amount of change in connection weight depends heavily on the current weight of the connection.  In particular, the HP memristor is insensitive to the effects of STDP where $W<0.1$, the PEO-PANI-like memristor is insensitive where $W>0.9$.  The linear memristor displays constant sensitivity.  Fig. 1(b) shows the effect of STDP on the weight in the more sensitive areas (for HP where $W>0.1$, for PEO-PANI where $W<0.9$), and compares them to the effect of STDP on the linear resistor over the same number of STDP steps.  In the case of the HP-like and PEO-PANI-like memristors, it can be seen that memristance can account for a change equal to 90\% of the total range of $W$ within 90 STDP steps; the same number of STDP steps gives a change in the linear resistor equal to 10\% of the total range of $W$.  

\section{Discovery Component}
\label{discovery}
In our GA, two parents are selected fitness-proportionately, mutated, and used to create two offspring.  We use only mutation to explore connection weight space; crossover is omitted as sufficient solution space exploration can be obtained via a combination of self-adaptive weight and topology mutations; a view that is reinforced in the literature, e.g. \cite{mu-only-needed}.  The offspring are inserted into the population and two networks with the lowest fitness deleted.  Parents stay in the population competing with their offspring. 

\subsection{Self-adaptive Mutation}
\label{sa-mutation}
We utilise self-adaptive mutation rates as in Evolution Strategies (ES) \cite{rechenberg}, to dynamically control the frequency and magnitude of mutation events taking place in each network. This allows for increased structural stability in highly fit networks whilst allowing less fit networks to search solution space more widely per GA application. Here, the $\mu$ $(0<\mu\leq1)$ value (originally $\sigma$, the rate of mutation per allele) of each network is initialized randomly uniformly in the range [0,0.25]. During a GA cycle, a parent’s $\mu$ value is modified as in (13), the offspring then adopts this new $\mu$, and mutates itself by this value, before being inserted into the population. The proportionality constant it set to 1 and thereore omitted.\\

$\mu \rightarrow \mu \exp^{N(0,1)}$ \hfill(13)\\

Only non-memristive networks can alter their connection weights via the GA.  Connection weights in this case are initially set during network creation, node addition, and connection addition randomly uniformly in the range [0,1].  Memristive network connections are always set to 0.5, and cannot be mutated from this value.  This forces the memristive networks to harness the plasticity of their connections during a trial to successfully solve the problem.

\section{Topology Mechanisms}
\label{topology}
In addition to self-adaptive mutation, we apply two topology alteration schemes to allow the modification of the spiking networks by adding/removing (i) hidden layer nodes (ii) neural connections.  This framework allows each network to control its own knowledge representation autonomously by adapting network topology to reflect the complexity of the problem considered.  All network types use these topology mechanisms.  Our self-adaptive topology mechanisms bare some resemblance to Takagi-Sugeno (TS) (neuro-) fuzzy models \cite{TS16,TS4,TS27} in that both parameter and self-organized structure learning occur (usually using a recursive least-squares algorithm for parameters and some rule density or utility metric for structure).  However TS systems  are commonly used for clustering and use multiple fuzzy rules to define a solution, rather than a single individual as in our case. 

\subsection{Neuroevolution}
\label{constructivism}
Given the nature of NC, it would be useful if appropriate network structure is allowed to develop until some task-dependent required level of computing power is attained.  A number of encoding variants have been developed specifically for neuroevolution, including Analog Genetic Encoding (AGE) \cite{4336123}, which allows for both neurons and connections to be modified,  amongst others, e.g. \cite{4016064}.  A popular framework is NeuroEvolution of Augmenting Topologies (NEAT) \cite{neat}, which combines neurons from a predetermined number of niches to encourage diverse neural utility and enforce niche-based evolutionary pressure.  This method has been shown to be amenable to real-time evolution ~\cite{1545941}.  Successful applications of neuroevolution range from real-world optimisation \cite{4079607} and classification \cite{5491156} to control \cite{1027752, 5508723}. 

In our system, each network has a varying number of hidden layer neurons (initially 9, and always $>$ 0).  Additional neurons can be added or removed from the single hidden layer based on  two new self-adaptive parameters, $\psi$ $(0<\psi\leq1)$ and $\omega$ $(0<\omega\leq1)$. Here, $\psi$ is the probability of performing neuron addition/removal and $\omega$ is the probability of adding a neuron; removal occurs with probability $1- \omega$. Both have initial values taken from a random uniform distribution, with ranges [0,0.5] for $\psi$ and [0,1] for $\omega$.  Offspring networks have their parents $\psi$ and $\omega$ values modified using (13) as with $\mu$, with neuron addition/removal taking part after mutation.  Added nodes are initially excitatory with 50\% probability, otherwise they are inhibitory.

\subsection{Connection Selection}
\label{conn-sel}
Feature selection is a way of streamlining input to a given process.  Automatic feature selection includes wrapper approaches (where feature subsets change during the running of the algorithm \cite{kohavi-john}) and filter approaches (where the subset selection is a pre-processing step \cite{koller-sahami}).  The connectivity pattern of artificial neural networks was first evolved by Dolan and Dyer \cite{dolan-dyer}.  A comparitive study can be found in \cite{neuroevo-arch-learn}.

In this paper we allow any connection to be individually enabled/disabled.  During a GA cycle a connection can be enabled or disabled based on a new self-adaptive parameter $\tau$ (which is initialized and self-adapted in the same manner as $\mu$ and $\psi$).  If a connection is enabled for a non-memristive network, its connection weight is randomly initialised uniformly in the range [0,1], memristive connections are always set to 0.5.  All connections are initially enabled for new networks.  During a node addition event, new connections are set probabilistically, with $P(connection\_enabled) = 0.5$.  Connection Selection is particularly important to the memristive networks. As they cannot alter connection weights via the GA, variance induced in network connectivity patterns plays a large role in the generation of useful STDP patterns.  In the context of NC, an evolutionary algorithm could conceivably tinker with connection structure as a means of homeostatic fault tolerance and recovery, as well as a compression technique to reduce the number of active synapses.

\section{Environmental Setup}
\label{environment}

Our chosen robotics simulator was Webots \cite{webots04}, a platform that is popular amongst the research community.  Alternatives are summarised in \cite{robot-sim-survey}.  Webots was selected due to the accuracy of its simulations and prevalence of successful applications in the literature.  Examples include evolution of simulated Khepera controllers to avoid obstacles \cite{evo-control-mob-rob}, showing the suitability of Webots to an evolutionary approach.  Tellez and Angulo \cite{tellez-ang} apply incremental neuroevolution to successfully generate complex behaviours from intitially-less-complex environments or sensory configurations.  Hierarchical neural control is exploited to guide a simulated Khepera around a T-maze using self-organising neural networks similar to our own \cite{paine-how-hierarchical}. 

\subsection{The Agent}
\label{agent}
The agent was a simulated Khepera II robot with 8 light sensors and 8 IR distance sensors (see Fig. 3(a)).  At each timestep (32ms in real time), the agent sampled its light sensors, whose values ranged from 8 (fully illuminated) to 500 (no light) and IR distance sensors, whose response values ranged from 0 (no object detected) to 1023 (object very close).  All sensor readings were scaled to the range [0,1] for computational reasons (0 being unactivated, 1 being highly activated).  Six sensors were used to comprise the input state for the SNN, three IR and three light sensors at positions 0, 2 and 5 as shown in Fig. 3(a).  Additionally, two bump sensors were added to the front-left and front-right of the agent to prevent it from becoming stuck against an object.  If either bump sensor was activated, an interrupt was sent causing the agent to reverse 10cm and the agent to be penalised by 10 timesteps.   Movement values and sensory update delays were constrained by Webots Khepera data.  Three actions were possible: forward, (maximum movement on both left and right wheels) and continuous turns to both the left and right (caused by halving the left/right motor outputs respectively).   Actions were calculated once at the end of each timestep from the output neuron classifications: ({\em high}, {\em high}) or ({\em low}, {\em low}) = forwards, ({\em high}, {\em low}) = left turn, ({\em low}, {\em high}) = right turn.

\subsection{The Environment}
\label{env}
The agent was located within a walled arena which it could not leave, with coordinates ranging from [-1,1] in both $x$ and $y$ directions and walls around the boundary having height $z=0.15$.   Adding to the complexity of the environment, a three-dimensional box was placed centrally in the arena, with vertices on ``ground level'' at ($x=-0.4$, $y=-0.4$), ($-0.4$, $0.4$), ($0.4$, $0.4$), and ($0.4$, $-0.4$), and raised to a height of $z=0.15$.  A light source, modelled on a 15 Watt bulb, was placed at the top-right hand corner of the arena ($x=1$, $y=1$).  The agent initially faces North, and its initial start position was constrained to the range $x+y < -1.5$.  The agent must traverse the environment and approach the light source to receive reward.  The environment is shown in Fig.3(b).

When the agent reached the goal state (where $x+y\ge1.6$), the responsible network received a constant fitness bonus of 2500, which was added to the fitness function $f$ outlined in (14).  The denominator in the equation expresses the difference between the position of the goal state (1.6) and the current agent position ($posx$ and $posy$), and $st$ is the number of timesteps taken to solve.  The minimum value of this function is capped so that $f > 0$.  The fitness of an agent is calculated at the end of every timestep, with the highest attained value of $f$ during the trial kept as the fitness value for that network.  Optimal performance gives $f=11800$, which corresponds to 700 timesteps from start to goal state with no collisions.\\

$f = (1/(1.6 - (|posx+posy|))) \times 1000 - st$\hfill(14)\\

\begin{figure}[!t]
\centering
	\subfloat[]{\label{highfit}\includegraphics[width=4cm,height=4cm]{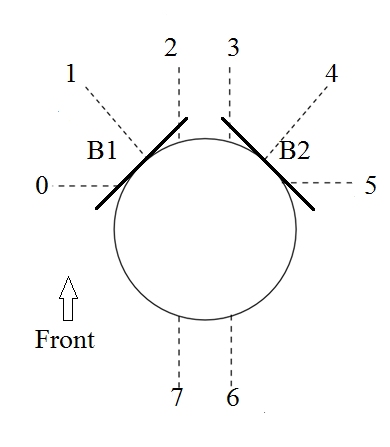}}     
	\subfloat[]{\label{avgfit}\includegraphics[width=4cm,height=4cm]{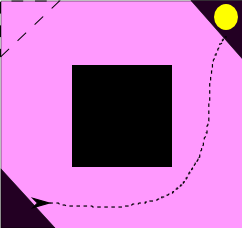}}
\caption{(a)Khepera sensory arrangement.  3 light sensors and 3 IR sensors share positions 0, 2 and 5.  Two bump sensors, B1 and B2, are shown attached at 45 degree angles to the front-left and front-right of the robot.(b)The test environment.  The agent begins in the lower-left and must reach a light source (circle) in the upper-right, circumnavigating the central obstacle.  An example agent path is shown (dotted line).  In the dynamic reward scenario (Section ~\ref{dyn}), the reward is moved to the upper-left.}
\label{fig_sim}
\end{figure}

\section{Experimental Setup}
\label{experimental-setup}
In the following experiments we gauged the impact of both types of memristive synapse, comparing to a benchmark systems containing (i) memory-augmented linear resistors (ii) constant-valued connections.  To aid clarity, we adopt a shorthand of ``PEO'' for networks containing only PEO-PANI connections.  Likewise, ``HP'', ``LIN'' and ``GA'' networks refer to networks containing only HP memristors, linear resistors, and constant connections respectively.\footnote{In traditional neural network terminology, the {\em objective} of the networks is to find a suitable sensor-motor mapping to allow for navigation, the {\em function} is the function that approximates this mapping, and the {\em compactness} is the minimal network topology as shown in the results sections.}

An experiment began with the generation of 100 networks of a given type (HP/PEO/LIN/GA).  Every network in the population was then trialed on the test problem, with a maximum of 4000 timesteps per trial (long enough to allow for initial exploration).  After this, 1000 generations of GA application took place and newly-generated networks were trialed on the test problem.  Every 20 generations, the current state of the system was observed and used to create the results that follow.  The entire process can be described as a {\em system}, with one system per connection type.  All experiments were repeated 30 times per system.  In any hardware implementation, the final solution would be the single fittest network from the population.

As the robot's start location was tightly constrained, we were able to compare system performance, defined as the first generation in which any network in that system found the goal state.  This measure produced 30 numbers, one per experimental repeat, that allowed us to perform t-tests to compare the respective goal-seeking performance of the four systems.  In Table~\ref{orig-ttest}, ``Performance'' was the average performance per network type as outlined above. ``High fitness'' refers to the mean fitness of the highest-fitness network in each run. ``Neurons'' were the average final connected neurons per network in the population  and ``Connectivity'' was the average percentage of enabled connections in the population.  Statistical signifance was assessed on the 5\% scale.

 SNN parameters were {\em initial hidden layer nodes}=9, $a=0.3$, $b=0.05$, $c=0.0$, $c\_ini=0.5$, $y_{thresh }= 1.0$ and $output \; window\; size=21$. Memristive parameters were $R_{on}=0.01$, $R_{off}=1$, $\beta=100$, $mem\_lifetime= 1000$, $last\_spike=3$, $spike\_threshold=4$.  During a trial the variable connections in the networks may alter their weights via STDP.  After every trial, variable connections were reset to their original weight of 0.5.  

\begin{figure}
\centering
	\subfloat[]{\label{highfit}\includegraphics[width=7cm,height=4.4cm]{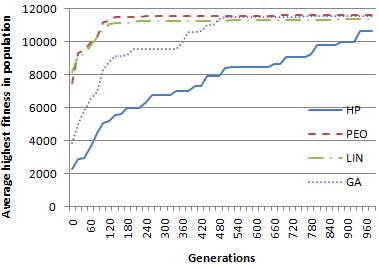}}             
	\subfloat[]{\label{avgfit}\includegraphics[width=7cm,height=4.4cm]{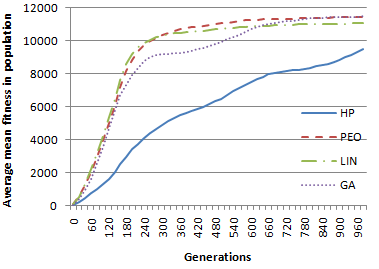}}  \\    
	\subfloat[]{\label{nodes}\includegraphics[width=7cm,height=4.4cm]{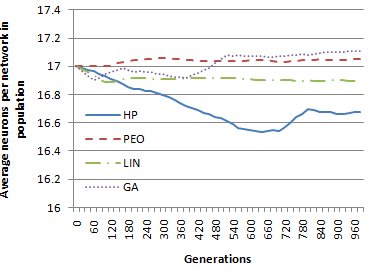}}       
	\subfloat[]{\label{conns}\includegraphics[width=7cm,height=4.4cm]{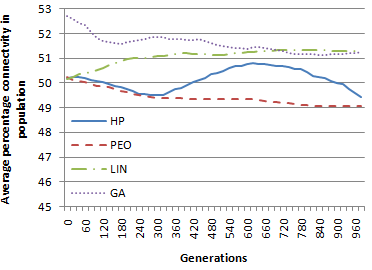}}  
\caption{(a) Average fitness of highest-fitness network per run (b) average fitness of entire population per run (c) average connected hidden layer nodes (d) average enabled connections for the first experiment}
\label{orig-perf}
\end{figure}

\begin{figure}
\centering
	\subfloat[]{\label{mu}\includegraphics[width=7cm,height=4.4cm]{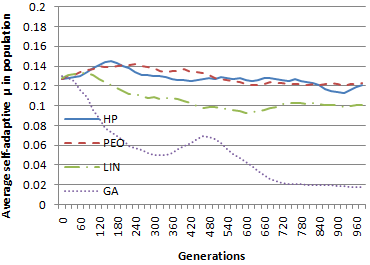}}               
	\subfloat[]{\label{psi}\includegraphics[width=7cm,height=4.4cm]{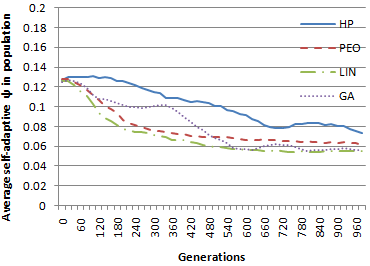}}      \\
	\subfloat[]{\label{omega}\includegraphics[width=7cm,height=4.4cm]{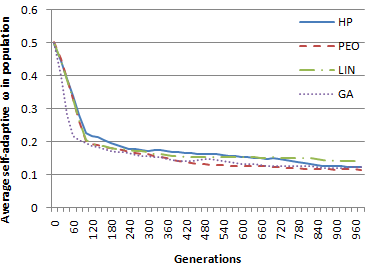}}       
	\subfloat[]{\label{tau}\includegraphics[width=7cm,height=4.4cm]{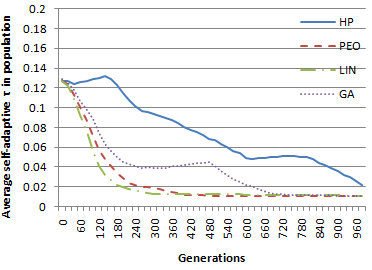}}  
\caption{(a) Average self-adaptive parameters (a) $\mu$  (b) $\psi$ (c) $\omega$ (d) $\tau$ in the population for the first experiment}
\label{orig-sa}
\end{figure}

\begin{table}
\begin{center}
\caption[]{Detailing t-test results (p values) for HP, PEO, LIN and GA systems on the test problem.}
\label{orig-ttest}
\begin{tabular}{|l|l|l|l|l|} 
\hline                &Performance       & High fitness      & Neurons    & Connectivity \\ 
\hline HP vs PEO   & {\bf 0.009}        &{\bf 0.033}   &  0.318       &0.983\\ 
\hline HP vs LIN    & {\bf 0.009}        & 0.106           &  0.601        &0.349\\ 
\hline HP vs GA     & {\bf 0.023}       & 0.091            & 0.699        &0.859\\
\hline PEO vs LIN  & 0.763                & {\bf 0.009}   & 0.130        &0.171\\
\hline PEO vs GA   &{\bf  0.027}       & {\bf 0.044}   & 0.107        &0.781\\
\hline LIN vs GA    & {\bf 0.019}       & 0.684            & 0.762        &0.289\\
\hline 
\end{tabular}
\end{center}
\end{table}

\subsection{Performance}
\label{homo-perf}
The most striking result from Table~\ref{orig-ttest} was that the PEO networks exceeded the GA networks statistically significantly, both in terms of performance (p=0.027) and high fitness (p=0.044).  LIN networks also outperformed GA networks (p=0.019), although did not have statistically better final fitness.  These results indicate that these networks learn to harness the plasticity of their connections alongside the topology variations introduced by connection selection to swiftly evolve goal-finding networks. 

In contrast, HP networks display significantly lower performance than all other network types (p=0.009 vs. PEO, p=0.009 vs. LIN, p=0.023 vs GA), as well as significantly worse high fitness than PEO networks (p=0.033).   This observed behaviour may be due to the memristance profile (see Fig.~\ref{stdpfig}(b)) being highly sensitive to the effects of STDP for high ($>0.9$) values of $W$, as well as being more likely to be stuck at low ($<0.1$) $W$ values, a notion echoed in \cite{neuro-asyn-stdp-memristive}.  It is reasoned that this combination of effects makes the memristor less suited to attaining highly-activated networks (network analysis reveals lower numbers of spikes per network, possibly preventing the network from reliably achieving certain output patterns). 

Figs.~\ref{orig-perf}(a) and~\ref{orig-perf}(b) show that the PEO and LIN networks share similar fitness profiles, both being distinctly quicker to attain high fitness values than HP and GA networks.  Both PEO and LIN networks solve the environment within 60 trials and attain their maximal fitness values within 150 trials.  GA networks reach lower final fitness values in a more gradual manner, reaching the maximal fitness value after 500 trials.  HP networks are slower still; highest fitness values are attained after 950 trials.  All systems eventually solved the problem, except 2 runs of HP networks.  A summary of averages and standard deviations is given in Table~\ref{orig-avg}.

\begin{table}
\begin{center}
\caption[]{Detailing averages and standard deviations for HP, PEO, LIN and GA systems on the test problem.}
\label{orig-avg}
\begin{tabular}{|l|l|l|l|l|} 
\hline                		&HP       	& PEO-PANI  & LIN   	 & GA 		 \\ 
\hline Perf                           & 526.1 (992.4)& 17 (34.6)& 14.7 (32.5)& 77.6 (130.0)\\
\hline  High fit 		 & 10660 (2280)       &11581 (303) 	&  11363  (398)     &11402  (277)	\\ 
\hline Avg fit    		&  9477 (3333)	 & 11454 (319)          &11058 (728)         &11420 (423)	\\ 
\hline Conns(\%)     	 & 49.42 (9.61)     	&  49.02  (4.63)        & 51.26 (4.06)       &51.19 (4.71)	\\
\hline Nodes 			 & 16.68 (1.74)       & 17.04 (0.09) &  16.89 (0.54)     &17.11 (0.57)	\\
\hline
\hline $\mu$  		 & 0.121  (0.09)     &0.123 (0.1)	& 0.115 (0.1)  	&  0.018 (0.01)   \\
\hline $\psi$   		& 0.073  (0.04)    	&0.062 (0.02)	& 0.019 (0.02)	&   0.056 (0.03)   \\
\hline $\omega$   		& 0.122 (0.11)     	&0.113 (0.07) 	& 0.135 (0.11)	&  0.122 (0.11)   \\
\hline $\tau$   		&0.022 (0.034)     	& 0.010 (0.01)	&0.010  (0.01)	&  0.011  (0.01)  \\

\hline 
\end{tabular}
\end{center}
\end{table}

\subsection{Topology}
\label{homo-top}
Although there were variations between the network types with regards to the numbers of hidden layer neurons, no statistically significant differences were observed (Table~\ref{orig-ttest} shows p-values ranging from p=0.107 to p=0.762).  While  PEO and LIN  networks show smooth profiles to their final average neuron numbers of 17.049 and 16.894 respectively (Fig.~\ref{orig-perf} (c)),  HP and GA networks show more unstable profiles to final numbers of 16.677 and 17.105 neurons respectively.   No significant differences were found with regards to connectivity, although Fig.~\ref{orig-perf} (d) shows a general order of PEO/GA networks being more densely connected, followed by HP networks and finally PEO networks.   Again,  profiles are more stable for PEO and LIN networks than they are for GA and HP networks.  

\begin{table}
\begin{center}
\caption[]{Detailing self-adaptation t-test results (p values) for all systems in the first experiment}
\label{orig-ttest-sa}
\begin{tabular}{|l|l|l|l|l|} 
\hline                &$\mu$            & $\psi$  & $\omega$  & $\tau$\\ 
\hline HP vs PEO   & 0.916            & 0.211   & 0.616        & 0.069  \\ 
\hline HP vs LIN    & 0.549   	  &{\bf 0.017}   & 0.226        & 0.07  \\ 
\hline HP vs GA     &{\bf  $<$0.001}   & 0.064   & 0.988        & 0.09  \\
\hline PEO vs LIN  & 0.618            & 0.178   & 0.243        & 0.525 \\
\hline PEO vs GA   & {\bf $<$0.001}   & 0.434   & 0.471        & 0.129  \\
\hline LIN vs GA    & {\bf $<$0.001}     & 0.471   & 0.465        & 0.458    \\
\hline 
\end{tabular}
\end{center}
\end{table}

\subsection{Self-adaptive Parameters}
\label{homo-sa}

In all cases, a lower parameter value is associated with a more stable evolutionary process, as such events are evolutionarily preferred to be less frequent within those networks.\\

\subsubsection{Mutation}
\label{homo-mut}
Being the only network to utilise the $\mu$ parameter, the GA networks final $\mu$ values were expectedly significantly different when compared to the other network types (all p-values $<$0.001) - Table~\ref{orig-ttest-sa}. Between the variable networks there were no statistically significant differences.  The GA mutation profile (Fig.~\ref{orig-sa} (a)) can be seen to rapidly increase from a value of 0.3 to 0.05 at generation 300, briefly climb to approximately 0.07 at generation 480, then descend smoothly to a final value of 0.019.  Other network $\mu$ profiles are irrelevent to the performance of those systems.\\

\subsubsection{Neuron addition/removal}
\label{homo-topology}
The probability of performing a neuron addition/removal event is encapsulated in the $\psi$ parameter.  One statistically significant difference can be seen in Table~\ref{orig-ttest-sa}, showing that such events are more likely to occur in HP networks than LIN networks.  This can be seen in  Fig.~\ref{orig-perf} (c) to drive the HP networks to lower numbers of neurons per network.  The probability of performing an addition rather than removal is governed by the $\omega$ parameter; Table~\ref{orig-ttest-sa} shows no statistically significant differences between the network types.   These results indicate that no single network type allows the evolutionary process to self-adapt to produce networks containing statistically fewer neurons whilst maintaining high performance.\\

\subsubsection{Connection addition/removal}
\label{homo-consel}
Connection selection is associated with the $\tau$ parameter.  Table~\ref{orig-ttest-sa} shows no statistically significant differences, although values comparing HP to other network types are all almost significant (p=0.069 vs. PEO, p=0.07 vs. LIN p=0.09 vs GA).  This difference is reflected in Fig.~\ref{orig-sa} (d), where HP networks produce the only initially upwards-trending profile which follows a markedly different curvature to the others.  Again, PEO and LIN network profiles can be observed to be similar to each other.

\section{Heterogeneous Mixtures of Memristors}
\label{hetero}
While the experiments presented in Section~\ref{experimental-setup} show the benefits of memristive connections, each networks behaviour under STDP is limited as homogeneous memristors follow identical STDP response curves as shown in Fig.~\ref{stdpfig}(a).  To increase the variety of plastic behaviours available to the network as a whole, we now extend the system to allow networks to be comprised of all three variable connection types (HP memristor, PEO-PANI memristor, and linear resistor).  As with our previous experiments, these networks should be replicable in hardware, provided that the myriad memristors can interface with a single neuron type, operate on the same scale, and possess similar electrical tolerances.  

Mixing different types of synaptic plasticity has been investigated previously \cite{soltoggio-het}\cite{maass-zador}\cite{flor-urz}.  In the first paper, interneural connections are affected by six distinct variations of the traditional Hebb rule.  In \cite{maass-zador},  spike transmission from synapse to neuron is probabilistic, with heterogeneous probabilities throughout the network.  Finally, \cite{flor-urz} uses four unique Hebbian learning rules for its connections; networks may be comprised of all four connection types.  All three papers consistently report that networks benefit from the inclusion of varied plasticity rules, mainly in terms of speed of goal finding, or encoding of functionality that is unattainable in homogeneously plastic networks.  Comparisons are made to GA,  PEO and LIN networks discussed in Section~\ref{experimental-setup}.  Experiments are conducted with the same parameters shown in Section~\ref{experimental-setup}, on the same environment as Section~\ref{environment}.

\subsection{Implementation}
\label{hetero-imp}
To facilitate the evolution of heterogeneous networks the system is altered in two regards, (i) connection creation and (ii) GA activity.\\

\subsubsection{Connection Creation}
On initialization of a new variable connection (via network creation, node addition, or connection addition), the type of that connection is selected probabilistically with $P=0.33$ of each type (HP-like memristor, PEO-PANI-like memristor, variable resistor) being selected.\\

\subsubsection{GA Activity}
Discovery is modified to allow one memristor type to mutate into another during a GA cycle.  As connections are always 0.5 before a trial begins and cannot be mutated,  $\mu$ has no role in the memristive networks.  Instead we use $\mu$ to control the rate of memristor type mutation taking place. During a GA cycle, after mutation, each connection in the child networks may alter to one of the two other connection types upon satisfaction of probability $\mu$.  Each network's value of $\mu$ is self-adapted as in equation (9), and is initially seeded randomly uniformly in the range [0,0.25] as with $\psi$ and $\tau$.   

\subsection{Performance}
\label{het-performance}
\begin{figure}
\centering
	\subfloat[]{\label{het-highfit}\includegraphics[width=7cm,height=4.4cm]{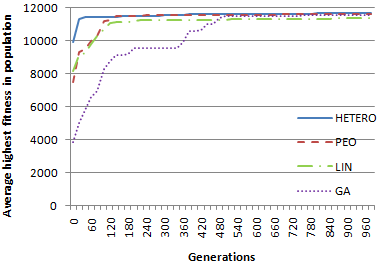}}                
	\subfloat[]{\label{het-meanfit}\includegraphics[width=7cm,height=4.4cm]{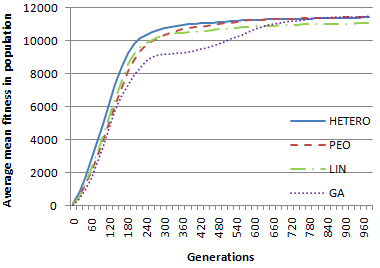}}      \\
	\subfloat[]{\label{het-nodes}\includegraphics[width=7cm,height=4.4cm]{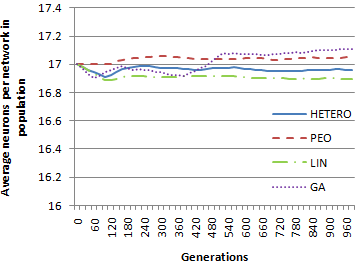}}       
	\subfloat[]{\label{het-newconns}\includegraphics[width=7cm,height=4.4cm]{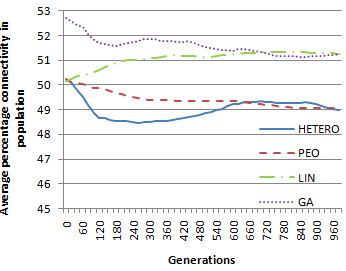}}  
\caption{(a) Average fitness of highest-fitness network per run (b) average fitness of entire population per run (c) average connected hidden layer nodes (d) average enabled connections in heterogeneous networks.}
\label{het-perf}
\end{figure}

Performance is shown in Table~\ref{het-ttest}, which reveals that heterogeneous networks have higher performance characteristics than PEO (p=0.026), LIN (p=0.043) and GA (p=0.003) networks.   Fig.~\ref{het-perf} (a) reveals that goal-finding behaviour is attained within 20 trials, faster than any homogenous network type.  The final ``high fitness'' value attained is higher than all other network types (Fig.~\ref{het-perf} (a)), and significantly higher than that of both LIN (p$<$0.001) and GA (p$<$0.001) networks (Table~\ref{het-ttest}).  Average fitness is shown in Fig.~\ref{het-perf} (b) and can be seen to attain near-optimal population-wide fitness after only 300 trials, an improvement over the other network types considered.  These results suggest that mixing synaptic behaviour allows the networks to more quickly attain higher performance characteristics. Averages and standard deviations are given in Table~\ref{het-avg}.

\begin{table}
\begin{center}
\caption[]{Detailing averages and standard deviations for HP, PEO, LIN and GA systems on the test problem.}
\label{het-avg}
\begin{tabular}{|l|l|l|l|l|} 
\hline                		&HET       	& PEO-PANI  & LIN   	 & GA 		 \\ 
\hline Perf			 &1.7 (4.8) 	& 17 (34.6)		& 14.7 (32.5)		& 77.6 (130.0)		\\
\hline  High fit 		 & 11696	 (186)      &11581 (303) 	&  11363  (398)     &11402  (277)	\\ 
\hline Avg fit    		&  11474	 (285)	 & 11454 (319)          &11058 (728)         &11420 (423)	\\ 
\hline Conns(\%)     	 & 48.97	(5.58)    	&  49.02  (4.63)        & 51.26 (4.06)       &51.19 (4.71)	\\
\hline Nodes 			 & 16.98	(0.65)	       & 17.04 (0.09) &  16.89 (0.54)     &17.11 (0.57)	\\
\hline
\hline $\mu$  		 & 0.074 (0.03)   &0.123 (0.1)	& 0.115 (0.1)  	&  0.018 (0.01)   \\
\hline $\psi$   		& 0.072 (0.02)   	&0.062 (0.03)	& 0.019 (0.02)	&   0.056 (0.03)   \\
\hline $\omega$   		& 0.132 (0.09)    	&0.113 (0.07) 	& 0.135 (0.11)	&  0.122 (0.11)   \\
\hline $\tau$   		&0.011 (0.01)     	& 0.010 (0.01)	&0.010  (0.01)	&  0.011  (0.01)  \\

\hline 
\end{tabular}
\end{center}
\end{table}

\subsection{Topology}
\label{het-topology}
As with the homogenous network comparisons, Table~\ref{het-ttest} reveals no significant differences with regards with final heterogeneous network neuron numbers.  Fig.\ref{het-perf} (c) visualises a steady profile that terminates slightly below its starting value of 17.  Percentage connectivity drops by approximately 1\% during the experiment to a final value of 49\%, shown in Fig.~\ref{het-perf} (d), giving a similar final value to PEO networks, lower than LIN and GA.  This is (just) significantly lower than LIN networks (p=0.049, Table~\ref{het-ttest}), although the actual difference is only 2\%.  Due to the general lack of statistical significance, it is demonstrated that the increased performance characteristics of heterogeneous networks are not offset by increased network complexity, and in some cases offer an improvement.

\subsection{Self-adaptive parameters}
\label{het-saparams}

\subsubsection{Mutation}
\label{het-mutation}

Heterogeneous and GA networks use $\mu$ to control different aspects of the GA cycle (HET networks use it to control the rate of switching of memristive behaviours, GA networks use it to alter connection weights).  Because of this, a statistically significant p-value $<$0.001 is seen between these network types (Table~\ref{het-ttest-sa}).  LIN and PEO networks do not use $\mu$ so comparisons are omitted.  Fig.~\ref{het-sa} (a) shows that the HET profile is more stable than the GA profile.\\  

\subsubsection{Neuron addition/removal}
\label{het-constructivism}

The profile of $\psi$ (the rate of constructivism events in the networks) is shown in Fig.~\ref{het-sa} (b) to descend to a final value of 0.072, higher than the other network types.  This is significantly higher than that of LIN (p=0.005) and GA (p=0.015) networks (Table~\ref{het-ttest-sa}), although this seems to correspond to heightened topology manipulation activity rather than different final neuron levels, as shown in Table~\ref{het-ttest}.  All network types show similar downward-trending profiles for the $\omega$ parameter, which is the probability of node additon as opposed to node removal upon satisfaction of $\psi$, shown in Fig.~\ref{het-sa} (c).  The similarity of these profiles is reflected in their respective p-values (Table~\ref{het-ttest-sa}), which show no significant differences.  These results indicate that the evolutionary process does not distinguish significantly between the variable connection types used in the networks.\\

\subsubsection{Connection addition/removal}
\label{het-consel}

\begin{figure}
\centering
	\subfloat[]{\label{het-mu}\includegraphics[width=7cm,height=4.4cm]{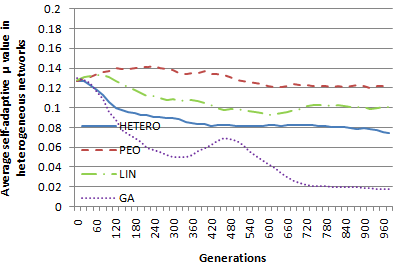}}                
	\subfloat[]{\label{het-psi}\includegraphics[width=7cm,height=4.4cm]{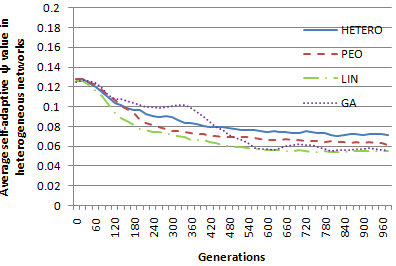}}      \\
	\subfloat[]{\label{het-omega}\includegraphics[width=7cm,height=4.4cm]{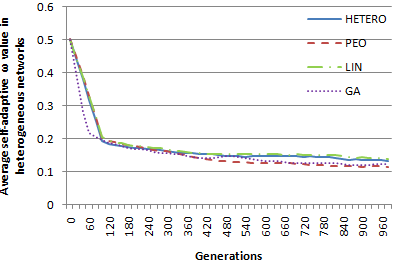}}       
	\subfloat[]{\label{het-tau}\includegraphics[width=7cm,height=4.4cm]{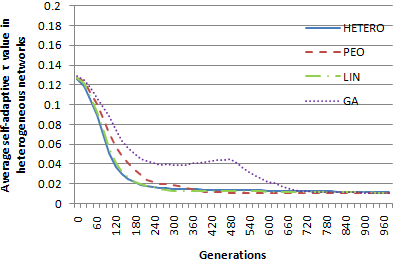}}  
\caption{(a) Average self-adaptive parameters (a) $\mu$  (b) $\psi$ (c) $\omega$ (d) $\tau$ in the population for heterogeneous networks}
\label{het-sa}
\end{figure}

Heterogeneous networks follow similar profiles to LIN and PEO networks, as visualised in Fig.~\ref{het-sa} (d).  Because of this, there are no statistical differences in terms of $\tau$.  Despite this lack of statistical significance in the controlling parameter, HET networks are significantly less connected than LIN networks (p=0.049, Table~\ref{het-ttest}.  This indicates that connection removal events are more likely to produce beneficial outcomes in HET networks than they are in LIN networks, as the frequency of those events is similar across the network types.\\

\begin{table}
\begin{center}
\caption[]{Detailing performance characteristic p-values when comparing heterogeneous memristive networks to PEO,LIN and GA networks}
\label{het-ttest}
\begin{tabular}{|l|l|l|l|l|} 
\hline HETERO vs.               &Performance         & High fitness 	& Neurons  	& Connectivity\\ 
\hline PEO	   		 &{\bf 0.026}  	 &  0.098  		& 0.718     	& 0.672  \\ 
\hline LIN			& {\bf 0.043}	&{\bf $<$0.001}	& 0.484	&{\bf 0.049}\\
\hline GA  			 &{\bf 0.003}           & {\bf $<$0.001} 	& 0.288       	& 0.537 \\
\hline 
\end{tabular}
\end{center}
\end{table}

\begin{table}
\begin{center}
\caption[]{Detailing self-adaptive parameter characteristic p-values when comparing heterogeneous memristive networks to PEO, LIN and GA networks}
\label{het-ttest-sa}
\begin{tabular}{|l|l|l|l|l|} 
\hline HETERO vs.               &$\mu$        	 & $\psi$ 		& $\omega$  & $\tau$\\ 
\hline PEO	   		 &{\bf 0.033}  	 & 0.136		& 0.184     	& 0.093  \\ 
\hline LIN			& 0.082		& {\bf 0.005}	&0.839	&0.302\\
\hline GA  			 &{\bf $<$0.001}  	 & {\bf 0.015}	& 0.495       	& 0.649 \\
\hline 
\end{tabular}
\end{center}
\end{table}

\section{Analysis of Heterogeneous Network Evolution}
\label{het-anal-top}
Although heterogeneous networks were found to have higher performance characteristics than all other network types,   it would be beneficial to know how STDP is used to benefit heterogeneous networks.  For example, are particular variable connection types more likely to be attached to excitatory or inhibitory neurons?  Is there an evolutionary preference to have a given type of variable connection attached to certain inputs, or driving a particular output neuron?  We focus on two broad themes; {\em evolution} (this section): evolutionary preferences to certain memristive configurations and {\em runtime} (section ~\ref{het-runtime}) - how STDP is used by those configurations to generate high-performance behaviour.

\begin{figure}[!t]
\centering
\includegraphics[width=3in]{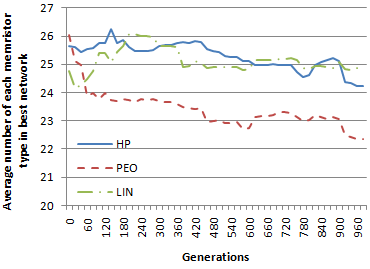}
\caption{Average number of each memristor type across the best network per run throughout the evolutionary process.}
\label{avg-memtypes}
\end{figure}

We average only the best network in each run, allowing us to focus on topological configurations that are beneficial.  Fig.~\ref{avg-memtypes} shows with HP and LIN components being preferred to PEO-PANI memristors.  Despite Table~\ref{top-best} showing no statistically significant differences, it is interesting to see the worst-performing memristor type from the homogenous networks (HP) being preferred to the best-performing (PEO-PANI).  This suggests that the evolutionary process finds a way to harness HP-like behaviour more readily when used in combination with other memristor types.

\begin{table}
\begin{center}
\caption[]{Detailing the distribution of memristor numbers by type as an average of the highest-performing network in each run.}
\label{top-best}
\begin{tabular}{|l|l|} 
\hline 					&P-value 	\\
\hline HP vs. PEO-PANI               	&    0.118	\\ 
\hline HP vs. LINEAR		&   0.609	\\
\hline PEO vs. LINEAR		&   0.054	\\	
\hline 
\end{tabular}
\end{center}
\end{table}

\subsection{Memristor Types per Layer}

\begin{figure}
\centering
	\subfloat[]{\label{mpl-IH}\includegraphics[width=8cm,height=5cm]{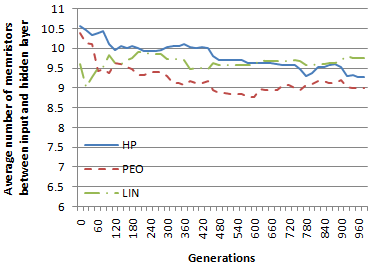}}     \\           
	\subfloat[]{\label{mpl-HH}\includegraphics[width=8cm,height=5cm]{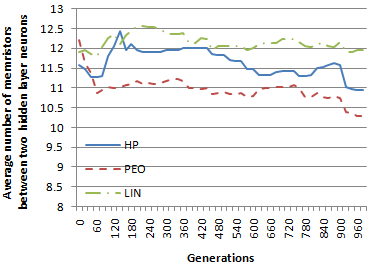}}      \\
	\subfloat[]{\label{mpl-HO}\includegraphics[width=8cm,height=5cm]{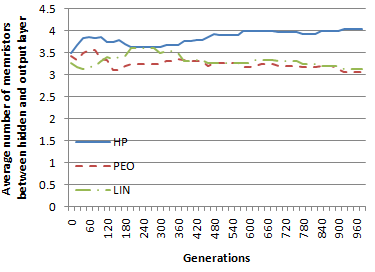}}       
\caption{Average number of each memristor type (a) between input and hidden layer (b) within hidden layer (c) between hidden and output layer in the best network per run through the evolutionary process.}
\label{memtype-per-layer}
\end{figure}

We now consider the specific positions of memristors in the networks.  As the networks consist of three layers, memristors can be classified based on the layers of the neurons that they connect, e.g. {\em input}, {\em hidden}, or {\em output}.  Figs.~\ref{memtype-per-layer} confirms results seen in  Fig.~\ref{avg-memtypes}, specifically that PEO-PANI memristors are universally more sparsely utilised than the other connection types.  In all cases, PEO-PANI memristors become the minority within 200 generations.

Two main significant results are shown in Table~\ref{top-layer}.  Firstly, LIN connections are preferred to both HP (p=0.045) and PEO-PANI (p=0.012) types when connecting two hidden layer neurons (Fig.~\ref{memtype-per-layer} (b)).  A feasible explanation is that the networks benefit from a basis of stable (e.g. linear) communications within the hidden (processing) layer to generate reliable action sequences.  More importantly, this result indicates that more linear memristors, if physically realised, could play an important role in future NC implementations.  Secondly, HP memristors are significantly (p=0.04) preferred to PEO-PANI memristors when connecting hidden neurons to output neurons;  Fig.~\ref{memtype-per-layer} (c) shows HP memristors are by far the most popular choice in this role.  HP memristors appear to be more suited to reliably reduce the number of spikes in the output trains to generate {\em low} output classifications when a turn is required. 

\begin{table}
\begin{center}
\caption[]{Detailing the distribution of memristor numbers by layer as an average of the highest-performing network in each run.}
\label{top-layer}
\begin{tabular}{|l|l|l|} 
\hline Memristor position	&Comparison			&P-value 	\\
\hline 	Input - hidden	&HP vs. PEO-PANI           	&    0.710	\\ 
\hline 				&HP vs. LINEAR		&   0.543	\\
\hline 				&PEO vs. LINEAR		&   0.339	\\	
\hline 	Hidden - hidden	&HP vs. PEO-PANI           	&    0.482	\\ 
\hline 				&HP vs. LINEAR		&   {\bf 0.045}	\\
\hline 				&PEO vs. LINEAR		&   {\bf 0.012}	\\	
\hline 	Hidden - Output	&HP vs. PEO-PANI           	&   {\bf 0.04}	\\ 
\hline 				&HP vs. LINEAR		&   0.079	\\
\hline 				&PEO vs. LINEAR		&   0.839	\\	
\hline 
\end{tabular}
\end{center}
\end{table}

\subsection{Position Analysis}

\begin{table}
\begin{center}
\caption[]{Detailing the distribution of memristor numbers relative to presynaptic and postsynaptic neuron types,  as an average the highest-performing network in each run.}
\label{top-neuron-type}
\begin{tabular}{|l|l|l|l|} 
\hline Neuron location	& Neuron type	&Comparison			&P-value 	\\
\hline Presynaptic		& Excitatory		&HP vs. PEO-PANI           	&    {\bf 0.018}	\\ 
\hline 				& 			&HP vs. LINEAR		&   0.805	\\
\hline 				&			&PEO vs. LINEAR		&   {\bf 0.015}	\\	
\hline 			 	& Inhibitory		&HP vs. PEO-PANI           	&    0.516	\\ 
\hline 				&			&HP vs. LINEAR		&   0.259	\\
\hline 				&			&PEO vs. LINEAR		&   0.874	\\	

\hline Postsynaptic		& Excitatory		&HP vs. PEO-PANI           	&    0.721	\\ 
\hline 				& 			&HP vs. LINEAR		&   0.368	\\
\hline 				&			&PEO vs. LINEAR		&   0.314	\\	
\hline 			 	& Inhibitory		&HP vs. PEO-PANI           	&    0.061	\\ 
\hline 				&			&HP vs. LINEAR		&   0.72	\\
\hline 				&			&PEO vs. LINEAR		&   0.183	\\
	
\hline 
\end{tabular}
\end{center}
\end{table}

Table~\ref{top-neuron-type} shows the relative numbers of memristor types that are connected (pre- or post-synaptic) to excitatory and inhibitory neurons.  PEO-PANI memristors are less preferred to the other connection types; p=0.018 vs. HP memristors and p=0.015 vs. LIN memristors, when an excitatory neuron is presynaptic.  As excitatory neurons are the sole method of activity generation, LIN may be preferred as they respond to STDP less dramatically, and can therefore more reliably maintain useful activity patterns.  Since excitatory spikes are also responsible for generating output spike trains, HP are reasoned to be preferred as they can reliably reduce network activity to generate {\em low} spike train classifications when required.  This supports the claim that certain memristor types are preferred in certain situations, and implies that the nonlinear activity of the HP and PEO-PANI memristors may be harnessed to alter, rather than preserve, network behaviour.    

\begin{table}
\begin{center}
\caption[]{Detailing the distribution of memristor numbers relative to input neuron function, as an average the highest-performing network in each run.}
\label{top-input-type}
\begin{tabular}{|l|l|l|} 
\hline Input source neuron	&Comparison			&P-value 	\\
\hline IR sensor		&HP vs. PEO-PANI           	&    {\bf0.033}	\\ 
\hline 				&HP vs. LINEAR		&   0.074	\\
\hline 				&PEO vs. LINEAR		&   0.933	\\	
\hline Light sensor		&HP vs. PEO-PANI           	&    1	\\ 
\hline 				&HP vs. LINEAR		&   0.379	\\
\hline 				&PEO vs. LINEAR		&   0.410	\\	
	
\hline 
\end{tabular}
\end{center}
\end{table}

Despite low general appearance rates within the networks, PEO-PANI memristors are significantly preferred to HP memristors when postsynaptic to an IR sensor (Table~\ref{top-input-type}, p=0.033).  As IR sensors respond only when near an obstacle, swift attainment of stable high activation to alter network acivation is required.  The PEO-PANI profile is also ideal to stabily create high-efficiacy connections via positive STDP which would make future obstacle avoidance response both stronger and quicker.  In contrast, no statistically significant values are found when the input neuron is attached to a light sensor.  Light sensors may be more ambivalent to more gradual synaptic efficiacy changes as the state space experienced by those sensors is less rugged than that experienced by the IR sensors; swift action perturbation is not required so any synapse type can function equally well.  

\section{Heterogenous STDP Analysis}
\label{het-runtime}
Synaptic plasticity acts to alter the influence of the connections on the activity of the network during a trial.  For this reason, we analyse the activity of the network as it solves the test problem, with particular attention paid to the role of STDP in behaviour generation. 

\begin{figure}
\centering
	\subfloat[]{\label{activation-all-weight}\includegraphics[width=7cm,height=4.4cm]{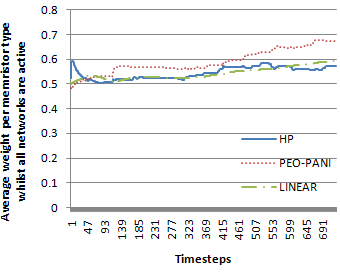}}           
	\subfloat[]{\label{activation-all-pos}\includegraphics[width=7cm,height=4.4cm]{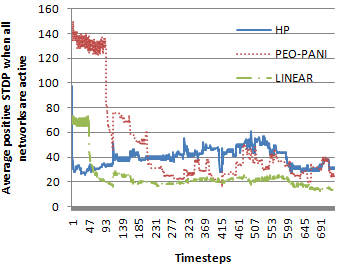}}    \\
	\subfloat[]{\label{activation-all-neg}\includegraphics[width=7cm,height=4.4cm]{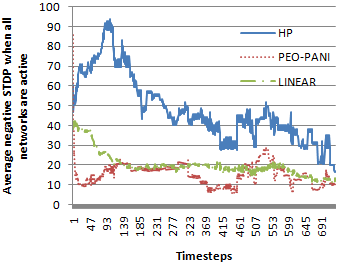}}     
	\subfloat[]{\label{best-het-net}\includegraphics[width=7cm,height=4.4cm]{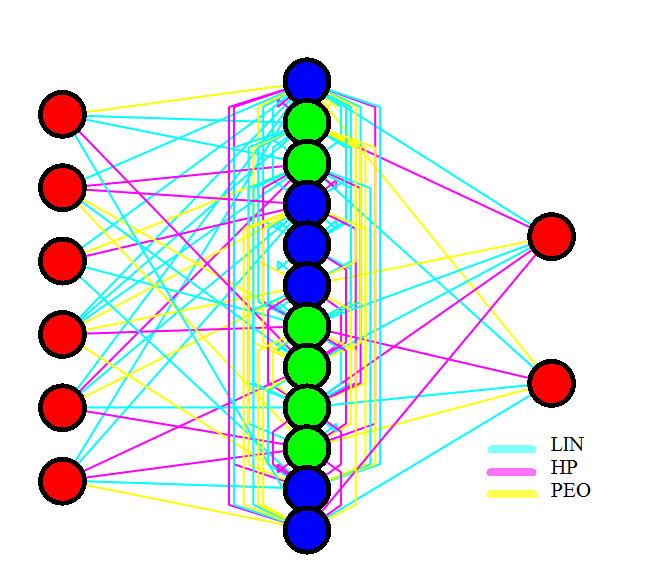}}    
\caption{Average (a) weight (b) positive STDP events (c) negative STDP events through activation, averaged over the best network from each run (d) Topology of the single best-performing network.  In the hidden layer, darker-coloured neurons are inhibitory and lighter-coloured neurons are excitatory.}
\label{activation-all}
\end{figure}

\subsection{Runtime analysis: Averages of Best Networks}

Each network took a differing amount of timesteps to solve the task.  Since averages are taken over the highest performing network in each run, there is an approximate correlation between the time that network executes the ``turn'', and the time it reaches the goal state and ends the trial.  Therefore, we are able to check for patterns within this timeframe.  Initial analysis revealed that the numbers of STDP events during a trial tend to osciallate between two or more values as the trial progresses.  To account for erronuous statistics that may arise as a result of these oscillations, all STDP values (and resultant weights) are averages over the previous 10 timesteps.

\subsubsection{Memristor weights}

Fig. ~\ref{activation-all} (a) shows the average weight per memristor type during runtime.  The PEO-PANI weight profile constantly increased throughout the duration of the trial, whereas the HP memristor and linear resistor weights terminated at approximately equal values.  PEO-PANI memristors were universally higher-weighted than HP memristors and linear resistors at the end of a trial; average final weights were HP memristor = 0.572, PEO-PANI memristor= 0.673 and linear resistor = 0.591.  Between HP and PEO-PANI memristors a statistically signifant p-value of 0.047 is observed, suggesting that PEO-PANI memristors act more like facilitating synapses than the other two memristor types.

\subsubsection{STDP}

STDP events were most prevalent during the first 250 timesteps  (Figs.~\ref{activation-all} (b) and ~\ref{activation-all} (c)).  Within this timeframe, HP memristors had low amounts of positive STDP (36.18) and high amounts of negative STDP (67.24). PEO-PANI memristors had high (84.45) positive STDP events, and low (16.48) negative STDP events.  Linear resistors possessed comparable amounts of both types of STDP; 31.23 positive and 25.04 negative events.   Because of this, HP memristors had significantly more negative STDP events than PEO-PANI did, and PEO-PANI underwent significantly more positive STDP events than HP (both p$<$0.001).  HP memristors experience significantly more negative STDP than negative STDP, with the inverse being true for PEO-PANI memristors (both p$<$0.001).  These results reinforce the view that the different memristor types are harnessed by the network as a result of being placed in favourable locations via evolution.

Following this period of heightened STDP activity, STDP events for all three variable connection types diminish and become more stable; networks use STDP to ``set up'' connection weights, which are then affected by further STDP to induce turning behaviour.  It should be noted that STDP generally involves more positive events than it does negative, e.g. the role of STDP is mainly to increase levels of activation within the networks.

\subsection{Turn Analysis of Best Overall Network}

We further refined the scope of our investigation to cover the single highest performing network,  shown in Fig.~\ref{activation-all} (d), which solved the task in 709 steps.  Motivation for focus on the turn is based on activity; as more STDP events occured during a turn this was an obvious timeframe to study.  

The network existed in a number of stable states oscillating between (usually 2) STDP values, which were observed throughout runtime. Turning motion began at timestep 293 and ended at timestep 372, during which periodic action switching behaviour between ``forward'' and ``right turn'' actions were observed.  Outside of this range, uniform ``forward'' actions were generated.  Performance characteristics of each memristor type at the turn event are shown in Table~\ref{runtime-stdp-memtype}.  HP memristors had lower rates of positive STDP with respect to negative STDP throughout the turn.  The opposite was true for PEO-PANI memristors, strengthening the notion that the two memristor types are evolutionarily preferred as facilitating and depressing synapses respectively.  During the turn, HP memristors were seen to maintain identical numbers of positive STDP events, whereas both PEO-PANI memristors and linear resistors kept identical negative STDP oscillations.

\begin{table}
\begin{center}
\caption[]{Detailing the STDP activity of memristor types during and after the turn event.}
\label{runtime-stdp-memtype}
\begin{tabular}{|l|l|l|l|} 
\hline 				& 	Memristor Type		&& \\
\hline 	 Start of turn		&	HP			&	PEO-PANI 		&  LINEAR\\
\hline Positive STDP		&	11 to 11/12		&	26/27 to 17/19	& 19/21 to 27		\\
\hline Negative STDP 	&	25/26 to 32/33	&	6/9 to14/15		& 43 to 25/29		\\
\hline&&& \\
\hline After turn& &&\\
\hline Positive STDP		&	11/12 to 13/17	& 37/41 to 58/63		&37/38 to 59/60		\\
\hline Negative STDP 	&	25/26 to 18/22	& 14/15 to 2/5		&25/29 to 22/23		\\
\hline 
\end{tabular}
\end{center}
\end{table}

In particular, two memristors were altered via STDP during runtime to achieve the desired behaviour, shown in Fig.~\ref{fig_sim}.    Firstly, the HP memristor connecting the second hidden node to the first output node underwent repeated negative STDP events, which due to the HP memristance profile enacted a swift decrease in conductivity.  This caused the initial turning motion by altering the spike train of the first output neuron from ``high activation'' to ``low activation''.  Correspondingly, the output action changed from constant forward motion to sequential ``forward'' and ``right turn'' actions.  Towards the end of the turn, this connection underwent a restrengthening due to different input node spike trains, allowing the first output neuron to achieve higher spiking frequency and produce a constant ``forward'' motion.  The second memristor in question was postsynaptic to the first input node and presynaptic to the 8th hidden node, which was strengthened at the end of the turn.  It was reasoned that this memristor allowed forwards motion to be generated by compensating for the change in light sensor values owing to the orientation of the agent changing.  Practically, the newly-strengthened memristor caused the 8th hidden neuron to spike more frequently.  As this neuron was connected to the first output neuron,  increased activity also caused the output neuron to spike more frequently, causing a ``high'' spike train classification which, in cooperation with the activity of the first memristor mentioned above, allowed for the generation of stable ``forward'' motion despite the new agent orientation. 

\begin{figure*}[!t]
\centerline{\subfloat[]{\includegraphics[width=5cm]{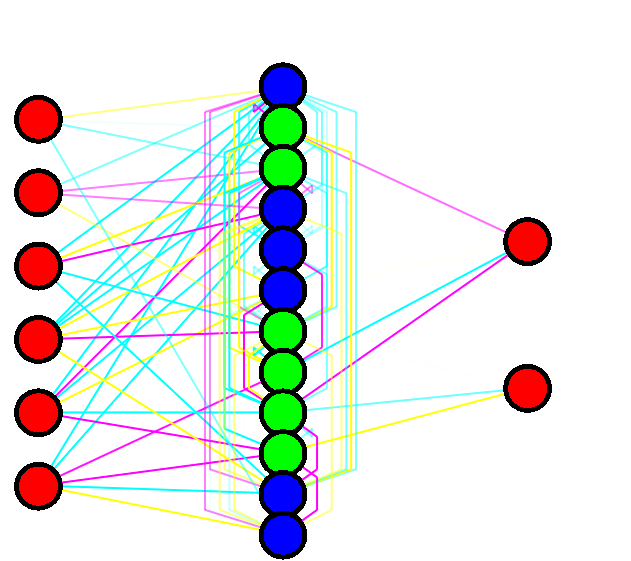}%
\label{pre-turn}}
\hfil
\subfloat[]{\includegraphics[width=5cm]{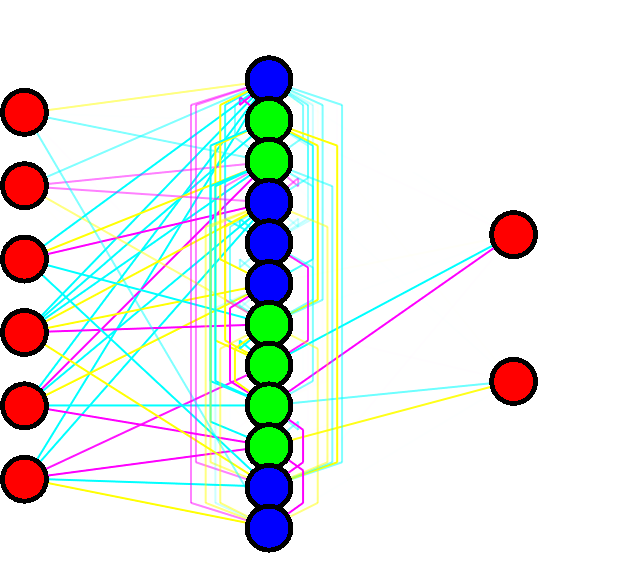}%
\label{during-turn}}
\hfil
\subfloat[]{\includegraphics[width=5cm]{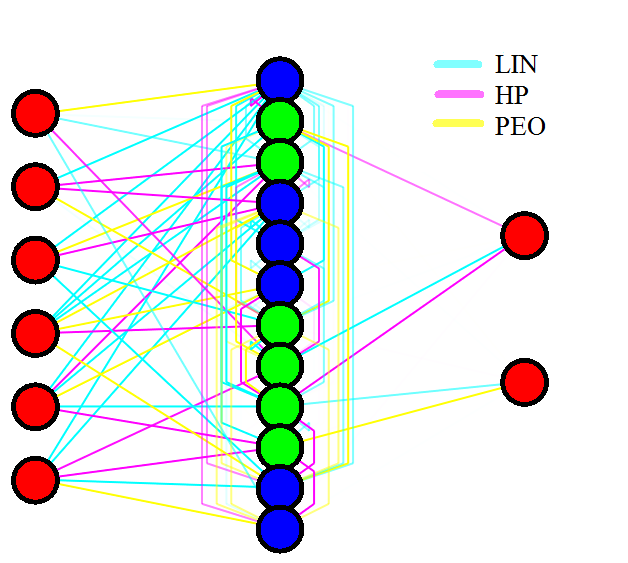}%
\label{after-turn}}}
\hfil
\caption{Showing the use of plasticity in the single best-performing network during activation, (a) pre-turn (b) during turn (c) after turn.  Darker coloured neurons are inhibitory and lighter coloured neurons are excitatory.}
\label{fig_sim}
\end{figure*}

\section{Dynamic reward scenario}
\label{dyn}
To further test the capabilities of the HET system, we ran an experiment based on the T-maze (e.g. ~\cite{tmaze}) scenario, where the agent must ``forget'' it's previously-learned behaviour after a time and adapt to a newly-positioned goal state.  Soltoggio \cite{solt-dyn} demonstrates the utility of plastic networks in such dynamic reward scenarios.

For continuity, the sensorimotor space was identical to the previous experiment (Fig.3), although the same adaptivity as in the T-maze is required.  We made the environment more challenging.  Firstly sensory noise was added based on Webots Khepera data; $\pm$2\% noise for IR sensors and $\pm$10\% noise for light sensors, all randomly sampled from a uniform distribution.  Wheel slippage was also included (10\% chance).  Secondly, the location of the reward changed from upper-right to upper-left during the lifetime of the agent (Fig.3(b)).  It should be noted that the light source does not move, e.g. with the reward in its second position, the agent is no longer performing phototaxis.  

Each trial was split into two parts, the reward was moved for part 2.  Membrane potentials and synaptic weights were not reset between these parts so that the agent had memory of the first part.  If the agent did not locate the goal in the first part, it cannot receive reward when the goal is moved.  A reward of 1 was given when the agent stabily found the first reward zone (part 1).  After this, the reward was relocated to the upper-left of the environment and part 2 commenced, continuing until the agent located the new reward zone (for a total fitness of 2), or the step limit was reached.  ``Performance'' was the number of trials the system took to find the second reward having located the first reward and was the main metric for comparison as it measured adaptation speed.   All other parameters are identical to those in Section~\ref{experimental-setup}.

In the following experiment, we compare the HET system to a benchmark GA system, and intend to demonstrate the utilty of memristive networks over those with static connections in this dynamic reward scenario.  Results, shown in Table~\ref{dynamic}, reveal that HET networks are universally preferable to GA networks, having higher performance, high fitness and average fitness, as well as lower connectivity and neuron numbers.  Significantly, HET networks are quicker at adapting to the change in reward location (p=0.037), suggesting that plastic memristive networks are suited to dynamic tasks.  Six of the GA networks could not locate both rewards.  The $\tau$ parameter was significantly lower (p=0.049), although this did not lead to a significant reduction in connectivity.  All other parameters ($\mu$, $\psi$, $\omega$) have different final values than those in the first experiment, demonstrating the context-sensitivity of the self-adaptation process. 

\subsection{Evolution Analysis}
Whereas the previous experiment saw HP being preferred to PEO-PANI when connecting hidden-output neurons, evolution now prefers both HP (p=0.007, avg 3.8) and PEO-PANI (p=0.047, avg 4.4) to LIN (avg 2.4) components.  This, coupled with the fact that LIN synapses are no longer significantly preferred to the other types between two hidden layer neurons, suggests that dynamic (nonlinear) activity is required by the networks to adapt rapidly.

HP memristors are also significantly preferred to PEO-PANI memristors (p=0.032) when connected to a light sensor, averages are 6 and 4.8 synapses of that respective type.  It is postulated that this favouring of an easily depressed connection type is one of the ways the networks evolve to deal with noise, which light sensors experience more than IR sensors.

\subsection{STDP Analysis}

\begin{figure}
\centering
	\subfloat[]{\label{activation-all-weight}\includegraphics[width=7cm,height=4.4cm]{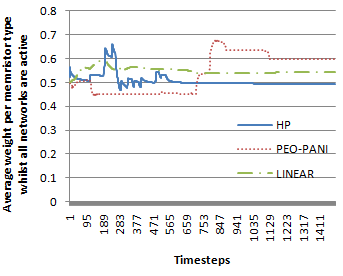}}                
	\subfloat[]{\label{activation-all-pos}\includegraphics[width=7cm,height=4.4cm]{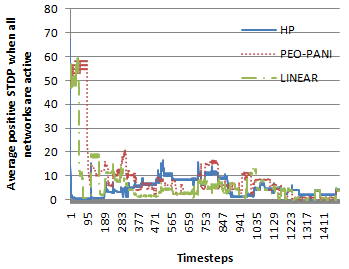}}    \\
	\subfloat[]{\label{activation-all-neg}\includegraphics[width=7cm,height=4.4cm]{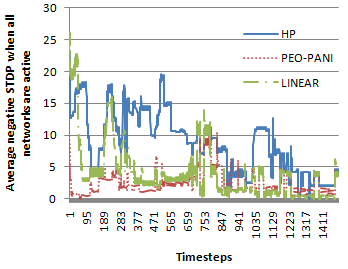}}      
	\subfloat[]{\label{activation-all-weight}\includegraphics[width=6cm,height=5cm]{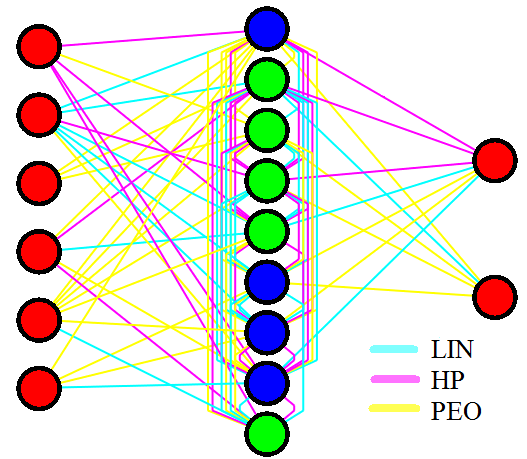}}     \\  
\caption{Average (a) weight (b) positive STDP events (c) negative STDP events through activation, averaged over the best network from each run in the dynamic scenario (d) highest-performing network topology.}
\label{dyn-all}
\end{figure}

Average synapse weights and STDP performance can be seen in Fig.~\ref{dyn-all}.  STDP is originally used as in the previous experiment, to ``set up'' connection weights.  The major result from STDP analysis of the best network from each run is that the average weight per synapse type varies between the two parts of the trial (Fig.~\ref{dyn-all}(a)).  In contrast to the previous experiment, PEO-PANI weights are low during the first part, and suddenly increase at the start of the second part (approximately timestep 750).  In all networks considered, PEO-PANI synapse weight varied more widely than the other types during a trial.  It is reasoned that PEO-PANI synapses were used as they can affect network activity more stabily, and more strongly per positive STDP event.  Average PEO-PANI weight (0.539) was significantly higher than average HP weight (0.507, p$<$0.001).

The use of STDP within the networks is shown in Figs.~\ref{dyn-all}(b) and(c).  STDP is intially similar to the previous experiment, with a slight increase in all types of positive STDP at timestep 750.  This coincides with a decrease in HP negative STDP, and increase in PEO and LIN negative STDP around the same timeframe.  Due to the nature of the PEO-PANI profile, this slight increase in positive STDP corresponds to the large increase in average weight seen in Fig.~\ref{dyn-all}(a), allowing the network to successfully solve the environment.  Overall, HP memristors undergo significantly more negative STDP than positive STDP, with the inverse being true of PEO-PANI memristors.  Noise is seen to be handled in some networks by STDP.  Specifically, light sensors are frequently seen attached to inhibitory neurons, which can act via STDP to reduce the impact of those inputs on the activity of the network.  To give some idea of topology, the best network is shown in Fig.~\ref{dyn-all}(d).  It should be noted that, despite the increased complexity of the task, this network contains less hidden layer neurons (17 vs. 20) and memristors (91 vs. 109) than the best network from the static environment.

\begin{table}
\begin{center}
\caption[]{Showing averages (standard deviations) and p-values of HET and GA systems on the dynamic reward scenario.}
\label{dynamic}
\begin{tabular}{|l|l|l|l|} 
\hline 				& 	HET			& GA			&	p-value  	\\
\hline 	Performance		&	57.8	(52.5)		& 541.4 (364)	&  	{\bf0.037}		\\
\hline 	High fit		&	2	(0)		& 1.8	(0.45)		&  	0.373		\\
\hline 	Avg fit		&	1.12	(0.08)		& 1.03 (0.04)	&  	0.155		\\
\hline 	Conns(\%)		&	52.21	(1.61)		& 52.78 (1.29)	&  	0.109		\\
\hline 	Nodes			&	16.79	(0.32)		&16.85 (0.12)	&  	0.739		\\
\hline  
\hline 	$\mu$			&	0.06	(0.01)		& 0.08 (0.02)	&  	0.111		\\
\hline 	$\psi$			&	0.08	(0.02)		& 0.09 (0.02)	&  	0.192		\\
\hline 	$\omega$		&	0.25	(0.09)		&0.30	(0.05)		&  	0.422		\\
\hline 	$\tau$		&	0.05	(0.01)		&0.06	(0.01)		&  	{\bf0.049}		\\
\hline 
\end{tabular}
\end{center}
\end{table}

\section{Summary}
In this paper we have demonstrated the first evolutionary approach to designing memristive SNNs for obstacle avoidance/dynamic reward tasks.  We have shown that plasticity can be harnessed by the networks via STDP to achieve more expedient goal-finding behaviour with no significant downside in terms of topological complexity.  Results indicate that, in possible NC implementations, heterogeneous mixtures of memristors possess advantages compared to both constant connections and networks of a single memristor type.  Self-adaptive parameters were found to alter dependent on the variable connection type in the network.  It is important to note that internal memristive network dynamics have {\em no analogue} in the GA case; those behaviours {\em cannot be replicated} by GA networks.  Overall it can be seen that all of our research questions have been answered and the hypotheses sufficiently demonstrated, as highlighted in the conclusions drawn from each set of experiments and summarised below.

The original hypothesis,  ``{\em that memristive synapses provide the networks with increased performance}'', was confirmed as PEO and LIN networks outperformed GA networks, and PEO networks evolved higher fitness solutions than GA networks.

The heterogeneous network hypothesis,  ``{\em to confirm that varied memristive behaviours can be harnessed by the evolutionary design process to provide further advantages, specifically that (i) certain functionality can be more easily achieved by certain memristor types (ii) combinations of memristor types are beneficial to the networks}'' was answered as heterogenous networks were higher performing than LIN, PEO and GA networks, and generated higher fitness solutions than LIN and GA networks.  They were also shown to outperform GA networks in a dynamic reward scenario.

Research question 1 -  ``{\em Does the evolutionary process allow for the successful generation of memristive networks that outperform constant-valued connections,  despite the memristors nonlinearity and given the potential for complex interactions within memristive networks?}'' - was answered as PEO networks were successfully evolved to outperform, and generate higher fitness solutions than, GA networks.

Reseach question 2 - ''{\em In the heterogeneous case, do mixtures of memristors provide better performance than other implementations?   How do such networks generate useful behaviour?}`` - was proven as heterogeneous networks had higher performance characteristics than PEO, LIN and GA networks. Useful behaviour was generated on an evolutionary level by assigning positions to the memristors based on their profiles, and on a runtime level by generating STDP to alter synaptic efficiacies to exploit properties of those profiles.

Research question 3 - ``{\em Is there an evolutionary preference in assigning specific roles to specific memristor types based on variations in their memristive behaviours?}'' was answered as a number of statistically significant differences with respect to the placement of specific connection types in certain positions in the networks were found.  

Biological brains contain mixtures of synapses that have specific types (e.g. depressing, facilitating) based on their performance characteristics \cite{347581}; results suggest that the evolutionary process casts the HP memristor as a depressing synapse and the PEO-PANI as an excitatory synapse within the heterogenous networks.  With statistical significance, in both static and dynamic scenarios, PEO-PANI synapses achieved higher average efficiency than HP synapses and underwent more positive STDP than negative STDP, as well as undergoing more positive STDP than HP memristors did.  The inverse is true of the HP memristor when compared to the PEO-PANI.  Biological brains also place these varied synapse types in certain contexts (e.g. ~\cite{t-and-d} gives examples of depressing synapses being typically found between two pyramidal neurons and facilitating synapses being frequently connecting between pyramidal and interneurons).  Initial findings provide compelling evidence that evolution of heterogeneous networks shares this feature; numerous examples have been reported herein including (i) PEO-PANI being preferred to HP when attached to IR sensors (ii) LIN being preferred to HP and PEO-PANI, being used to generate more stable behaviour between two hidden layer neurons (iii) HP being preferred to PEO-PANI when connecting to output layer neurons.  In the dynamic case, (i) HP were preferred to PEO-PANI when attached to light sensors (ii) PEO-PANI were preferred to HP and LIN when connecting to output neurons.  It is clear that the memristors are assigned types, and thus roles,  within the networks based on their profiles.  It is also shown that the role for a specific type can vary based on the environment the controller encounters.

The introduction highlighted the implementation of neuromorphic structures as motivation for conducting this research.  The use of physical equations to model memristive behaviour makes a future hardware implementation more viable.  Performance of the yet-to-be-realised LIN component indicates that more linear memristive behaviours may be beneficial, especially in a heterogeneous scenario.  As the memristors have a constant initial weight that is not affected by GA activity, memristive networks are initially handicapped with less degrees of behavioural freedom.  Despite this fact, they are shown to adapt by allowing environmental signals to alter synaptic efficiency to outperform (in some cases) the GA approach on the test problem.  The networks evolved are admittedly at a much smaller scale than those required by the neuromorphic paradigm.  Scalability is more likely to be possible due to the inclusion of constructivism and self-adaptive search parameters, provided that the innate self-organising properties of the networks can account for the increased complexity of intra-network communications.

\bibliographystyle{IEEEtran}

\bibliography{TEC-mems}

\end{document}